\documentclass[number,sort&compress]{elsarticle}

\usepackage{graphicx}
\usepackage{amsmath}   
\usepackage{lineno}
\nolinenumbers

\begin{document}

\begin{frontmatter}

\title{Status of the TORCH time-of-flight project}

\author[add0]{N.~Harnew\corref{cor}}
\ead{Neville.Harnew@physics.ox.ac.uk}
\author[add1,add2]{S.~Bhasin}
\author[add5]{T.~Blake}
\author[add1]{N.H.~Brook}
\author[add6]{T.~Conneely}
\author[add2]{D.~Cussans}
\author[add3]{M.~van~Dijk}
\author[add3]{R.~Forty}
\author[add3]{C.~Frei}
\author[add4]{E.P.M.~Gabriel}
\author[add0]{R.~Gao}
\author[add5]{T.J.~Gershon}
\author[add3]{T.~Gys}
\author[add0]{T.~Hadavizadeh}
\author[add0]{T.H.~Hancock}
\author[add5]{M.~Kreps}
\author[add6]{J.~Milnes}
\author[add3]{D.~Piedigrossi}
\author[add2]{J.~Rademacker}

\cortext[cor]{Corresponding author}

\address[add0]{Denys Wilkinson Laboratory, University of Oxford, Keble Road, Oxford OX1 3RH, United Kingdom}
\address[add1]{University of Bath, Claverton Down, Bath BA2 7AY, United Kingdom.}
\address[add2]{H.H. Wills Physics Laboratory, University of Bristol, Tyndall Avenue, Bristol BS8 1TL, United Kingdom}
\address[add3]{European Organisation for Nuclear Research (CERN), CH-1211 Geneva 23, Switzerland}
\address[add4]{School of Physics and Astronomy, University of Edinburgh, James Clerk Maxwell Building, Edinburgh EH9 3FD, United Kingdom }
\address[add5]{Department of Physics, University of Warwick, Coventry, CV4 7AL, United Kingdom}
\address[add6]{Photek Ltd., 26 Castleham Road, St Leonards on Sea, East Sussex, TN38 9NS, United Kingdom}

\begin{abstract}
TORCH  is a time-of-flight detector, designed 
to provide  charged $\pi /K$ particle identification  up to a momentum of 10\,GeV/$c$ for  a 10\,m flight path. 
To achieve this level of  performance, a time resolution of 15\,ps per incident particle is required.  
TORCH uses a plane of quartz of 1\,cm thickness as a source of Cherenkov photons, which are then focussed
onto square   Micro-Channel Plate Photomultipliers (MCP-PMTs) of active area 53 $\times$ 53\,mm$^2$, 
segmented into 8 $\times$ 128 pixels equivalent. 
A small-scale TORCH demonstrator with a customised MCP-PMT and associated readout electronics 
has been  successfully operated in a 5\,GeV/$c$ mixed pion/proton beam at the CERN PS facility.  
Preliminary results indicate that a single-photon resolution better than 100\,ps can be achieved. 
The expected performance of a full-scale TORCH detector for the Upgrade II of the LHCb experiment is also  discussed.

\end{abstract}


\end{frontmatter}

\newpage

\section{Introduction}

\begin{figure}[h]
\centering
\includegraphics[width=0.95\linewidth]{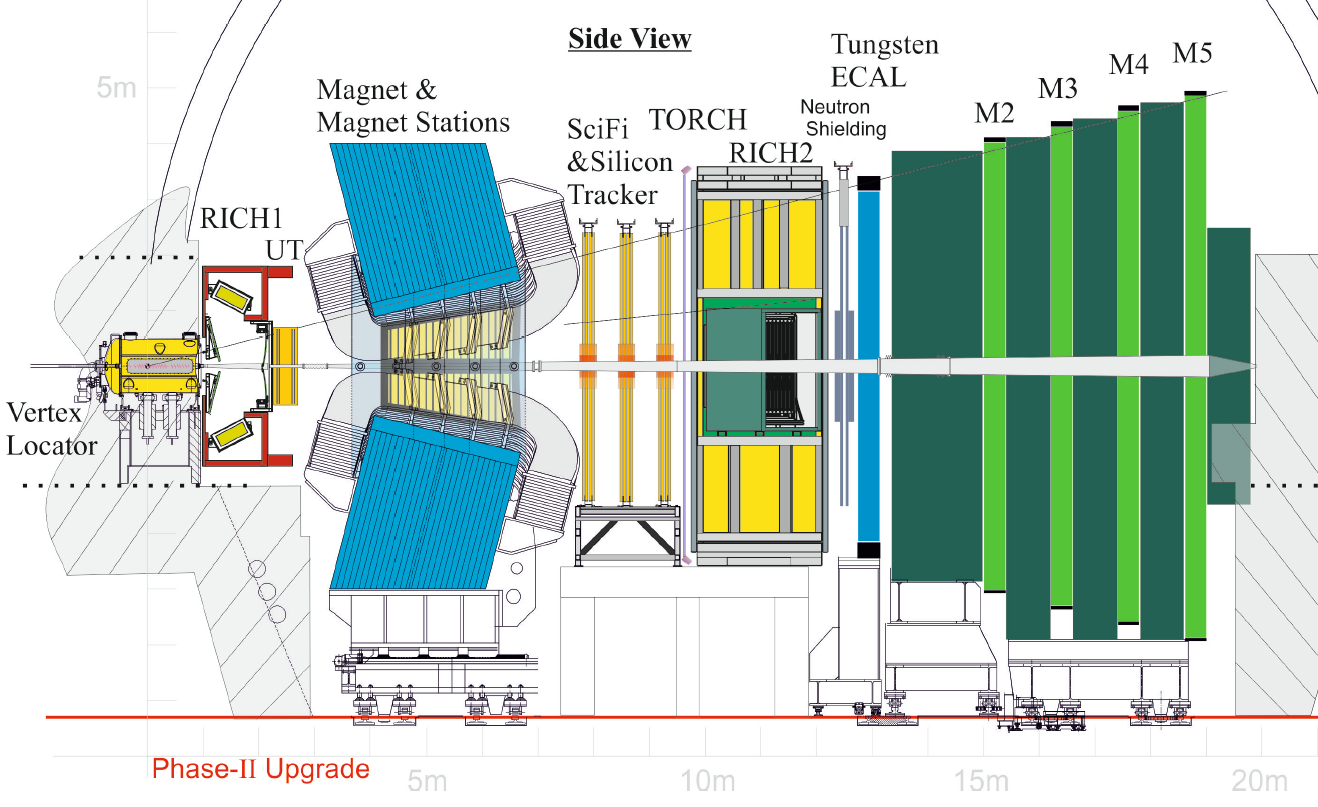}
\caption{A schematic of the LHCb experiment, showing TORCH located directly upstream of the
 RICH\,2 detector.}
\label{fig:LHCb-schematic}
\end{figure}

The TORCH (Time Of internally Reflected CHerenkov light) detector will measure the
 time-of-flight (ToF) 
of charged particles over large areas, with the aim  
to provide  Particle IDentification (PID) of pions, kaons and protons  
 up to 10 GeV/$c$ momentum and beyond\,\cite{Charles:TORCH}.
The difference in ToF between pions and kaons over a $\sim$10\,m flight path at 10\,GeV/$c$ is 35\,ps, 
hence to achieve positive identification of kaons, TORCH aims for a time resolution of
$\sim$10-15\,ps per track.
A specific application of TORCH  is for the 
LHCb Upgrade II experiment, where the detector would occupy an area of
30\,m$^2$ in front of the current RICH\,2 detector\,\cite{LHCb_PhaseII_upgrade}. The 
proposed experimental arrangement is shown in Fig.\,\ref{fig:LHCb-schematic}.

TORCH combines timing measurements with DIRC-style reconstruction, a technique pioneered by the  
BaBar DIRC\,\cite{ref:Babar} and  Belle\,II TOP\,\cite{ref:BelleII-TOP} collaborations.
The production of Cherenkov light  is prompt, hence TORCH uses
 planes of 1\,cm thick quartz as a source of fast signal, which also facilitates a modular design. 
Cherenkov photons travel to the periphery of the quartz plates by total internal reflection where they are 
reflected by a cylindrical mirror surface of a quartz block. This focuses the photons onto a plane of pixellated Micro-Channel Plate Photomultipliers  (MCP-PMTs)
where their positions and arrival times are measured.
The expectation is that typically 30 photons will be detected
per charged track, hence the required ToF resolution dictates the timing 
of single photons to a precision of around 70\,ps.

\begin{figure}
\centering
\includegraphics[width=0.41\linewidth]{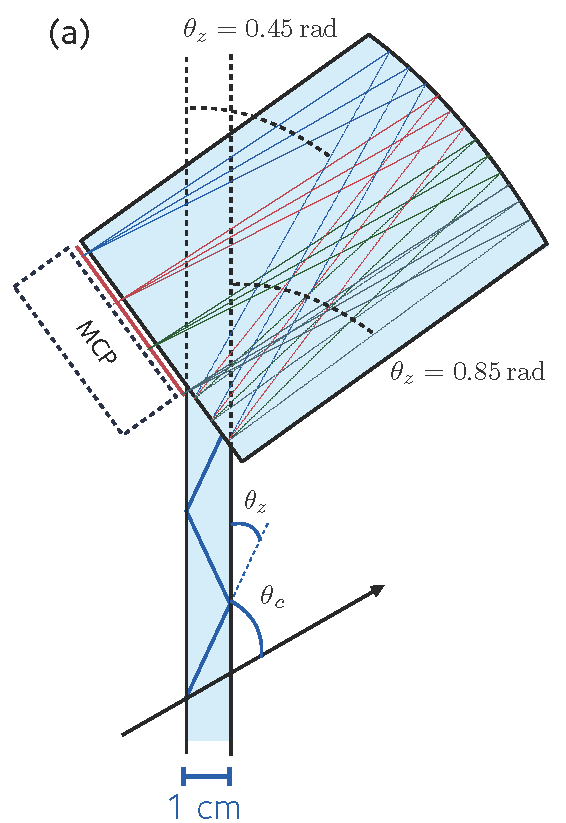}
\hspace{0.7truecm}
\includegraphics[width=0.32\linewidth]{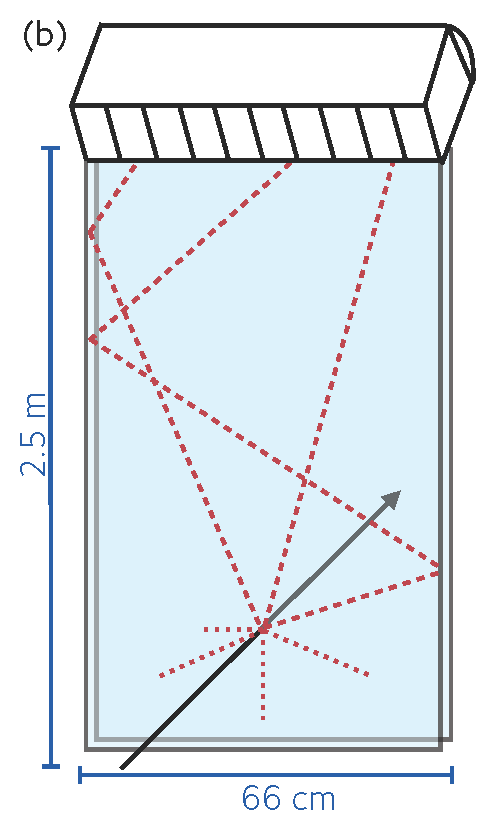}
\caption{Schematics  of a TORCH module showing possible reflection paths: 
(a) the focussing block and MCP-PMT plane, (b) a single  LHCb module.}
\label{fig:TORCH-schematics}
\end{figure}

A schematic of the TORCH geometry in the longitudinal and transverse  planes
is shown in Fig.\,\ref{fig:TORCH-schematics}, showing the focussing 
block and the LHCb modular arrangement.
For every photon hit in the MCP-PMTs,
the Cherenkov angle $\theta_c$\\

\noindent and the photon path length of propogation 
through  the quartz   $L$  is measured. From knowledge of the dispersion relation
within the quartz, a correction for  chromatic dispersion
is then made to the photon time of propogation. From simulation, a $\sim$1\,mrad precision is required 
on the measurement of the angles in both planes to achieve the required intrinsic timing 
resolution\,\cite{Charles:TORCH}. 

\section{MCP-PMT development}

In order to achieve the desired 1\,mrad angular precision, TORCH 
requires a photon detector with a fine granularity 
in the focussing direction and a coarse granularity in the non-focussing direction. 
The  pixel structure of MCP-PMTs can in principle be adjusted 
to the resolution required provided the charge footprint is small enough.
An anode granulatity of $128 \times 8$ pixels is chosen 
with a $53 \times 53$\,mm$^2$ active area on a 60\,mm pitch. 
MCP-PMTs are well known for fast timing of single photon signals, 
$\sim$30\,ps, however tube lifetime has been an issue in the past. 

\begin{figure}
\centering
\includegraphics[width=0.8\linewidth]{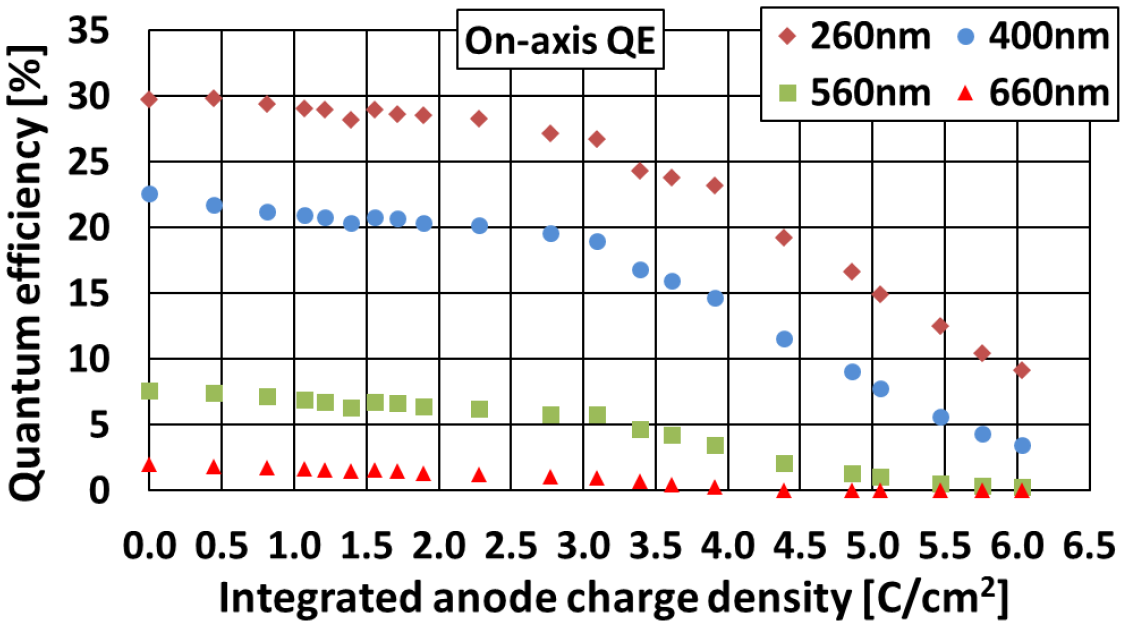}
\caption{The quantum efficiency of a Photek Phase\,1 MCP-PMT as
a function of wavelength and collected integrated charge, as measured on the tube axis.}
\label{fig:TORCH-ageing}
\end{figure}

A major  focus of the TORCH project has been on the  development 
of the MCP-PMTs, which has been conducted in collaboration with an industrial partner,  Photek (UK).
A three-phase R\&D programme was defined.  Phase\,1 saw the development of a single-channel 
MCP-PMT with extended lifetime,  accomplished with an
atomic-layer deposition (ALD) coating of the MCPs\,\cite{ref:Conneely}, and where an excellent timing resolution of better than  $\sim$35\,ps was also
achieved\,\cite{ref:Gys}.    The extended lifetime is required for the harsh environment of the LHC,
where integrated anode charges of at least 5\,Ccm$^{-2}$ are expected.
Lifetime measurements have since been conducted over a period of 2.5 years
using single-photon illumination from a blue LED, and are now 
reaching an integrated charge of 6.16\,Ccm$^{-2}$. 
Figure\,\ref{fig:TORCH-ageing} demonstrates the results. Some loss of quantum efficiency is seen 
above 3\,Ccm$^{-2}$, with a factor ~2 loss at 5\,Ccm$^{-2}$.  
A gain drop is also measured, but which can be recovered by an increase of MCP high 
voltage\,\cite{Gys_2016_RICH_proceedings}.
The MCP-PMT lifetime is close to the required performance, and it is expected that the
Phase\,3 tubes will improve on this.

The Phase\,2 MCP-PMTs are circular in construction with a 40\,mm diameter 
and a square 26.5 $\times$ 26.5\,mm$^2$  active area 
containing  4 $\times$ 32 pixels, a quarter size of the final geometry. 
Beam tests on these tubes were successfully completed in 2015/16  and are reported 
elsewhere\,\cite{Brook-2018}.  

The final TORCH devices, the Phase\,3 tubes,  have
high active area ($>$80\%), the required granularity, ALD coating to give the extended lifetime,  and
excellent time resolution.   
Ten Phase\,3 MCP-PMTs have been delivered from Photek and are currently under test.
The MCP-PMT has a square 53 $\times$ 53\,mm$^2$ active area with 64 $\times$ 64 pixels. 
The effective resolution of 128 $\times$ 8 pixels is achieved in the focussing and non-focussing directions by exploiting charge 
sharing between pixels\,\cite{ref:Castillo} and by ganging 
electronically 64 pixels into 8  (or 4), respectively.  
The MCP-PMT anode is connected to  a PCB  via Anisotropic Conductive Film (ACF) onto which a
readout connector is mounted.
The readout configuration  of the first  evaluated tube  has 64 $\times$ 4 pixels, compatible with a previous version of electronics,  
and used in the test-beam of November 2017. The second
configuration gives  64 $\times$ 8 pixels per tube for use with new electronics in the test-beam of 
June 2018\,\cite{Gao-2016-TWEPP2015-proceedings}.

\section{Beam tests with a small-scale TORCH demonstrator}

Several test-beam campaigns have been conducted 
between 2015--2018 at the CERN PS T9 beamline with a 5 GeV/$c$ mixed pion/proton  beam.
The  small-scale TORCH demonstrator\,\cite{Brook-2018} consists of a $12 \times 35 \times 1$\, cm$^3$ quartz radiator plate  with
a matching focusing block, both manufactured by Schott, Germany.
The radiator plate is mounted in an almost vertical position, tilted backwards by 5$^\circ$ with respect to the horizontal  incidence of the beam.

We report here preliminary results from the November 2017 campaign, where the demonstrator is
read out with a single Phase\,3 MCP-PMT in the 64 $\times$ 4 pixel configuration. The 
customised electronics system\,\cite{Gao-2016-TWEPP2015-proceedings} uses the 
NINO32\,\cite{Despeisse-2011-NINO32} and HPTDC\,\cite{Christiansen:HPTDC} chipsets, 
developed for fast timing applications of the ALICE experiment.
Proton-pion selection is achieved independent of TORCH using a pair of upstream Cherenkov counters, and 
two borosilicate finger counters (T1 and T2) separated by  a $\sim$11\,m flight path provide a time reference.

\begin{figure}[h]
\centering
\includegraphics[width=0.49\linewidth]{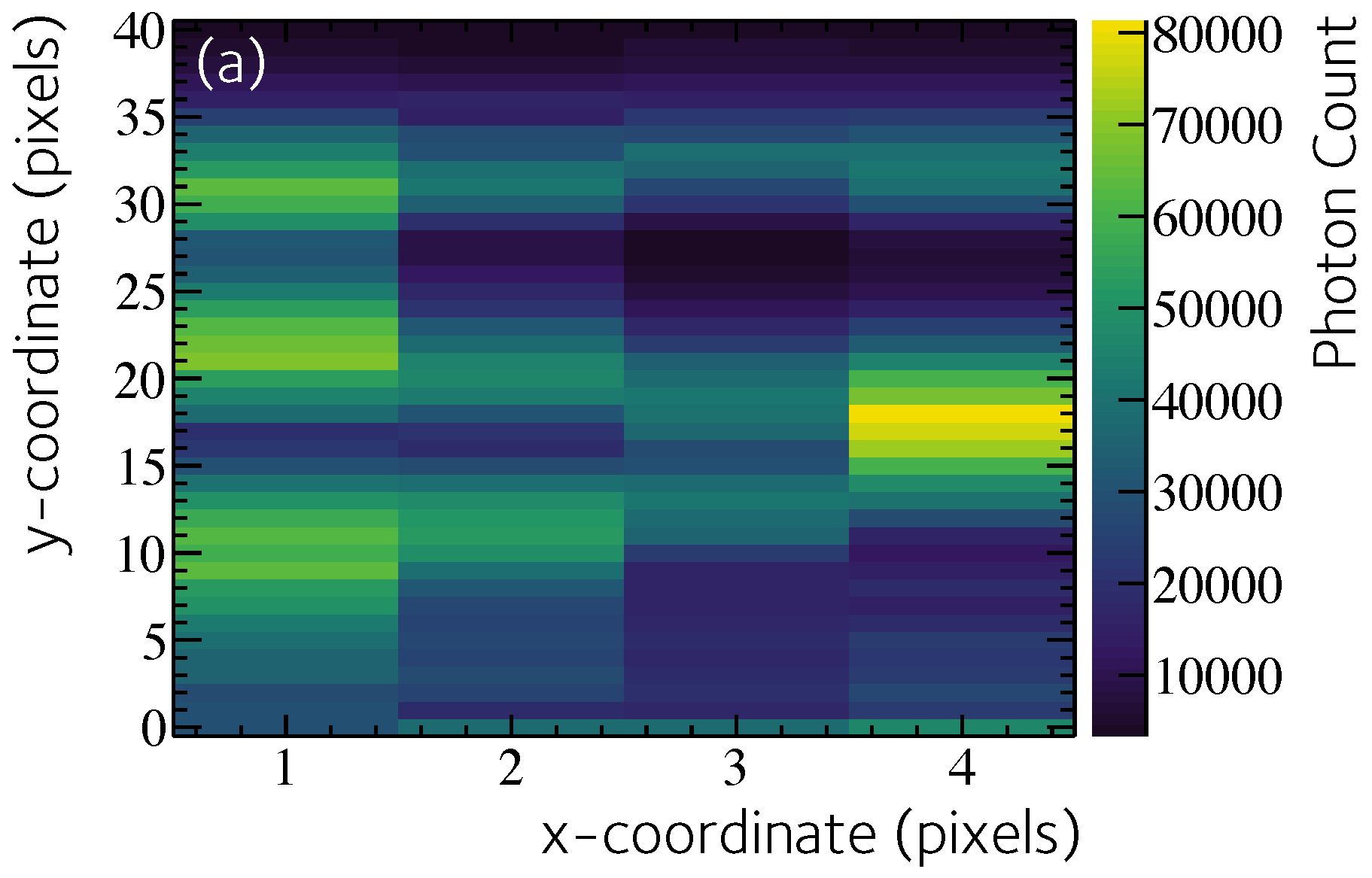}
\includegraphics[width=0.49\linewidth]{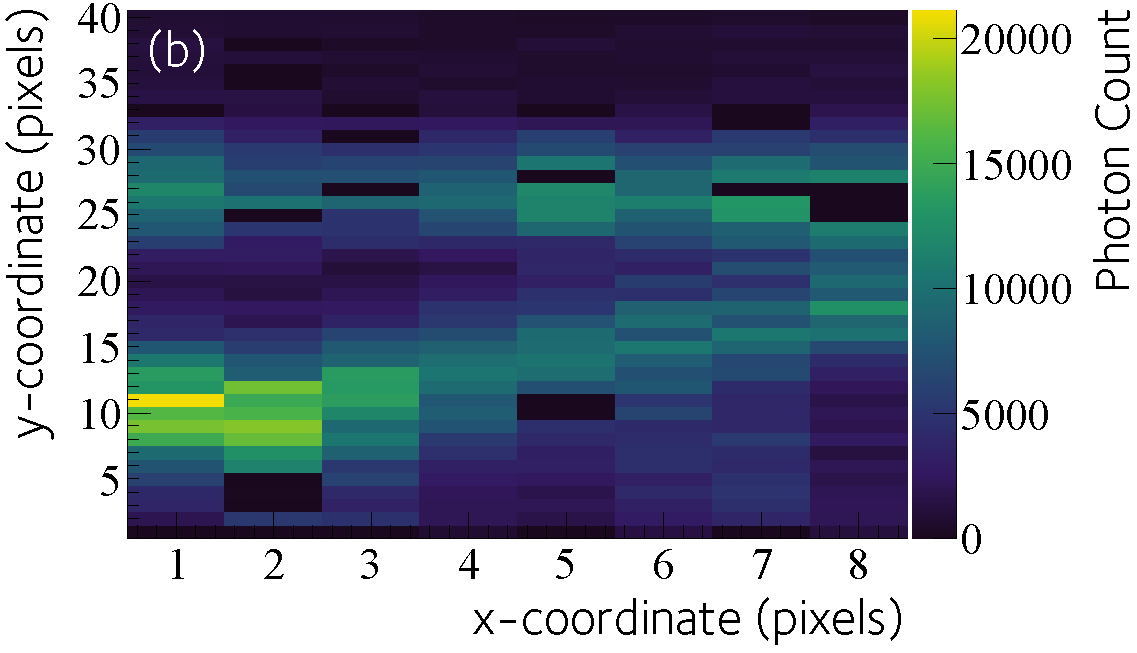}
\caption{The patterns of hits measured in  the TORCH demonstrator with 5\,GeV/$c$ pions.
(a)   64 $\times$ 4  pixel readout in November 2017 with the beam 14\,cm below the quartz radiator centre-line and the MCP-PMT located in its nominal position on the focal plane, and
 (b)  64 $\times$ 8 pixel readout in June 2018 with the beam at the quartz  centre-line and the MCP-PMT shifted 5\,mm upwards
with respect to configuration (a).}
\label{fig:hits}
\end{figure}

The pattern of measured  MCP-PMT hits for 5\,GeV/$c$ pions is shown  in Fig.\ref{fig:hits}\,(a). Here 
clustering has been  applied over simultaneous MCP-PMT column hits to obtain the centroid position of each photon.
The beam impinges approximately 
14\,cm below the plate centre-line and close to the plate side, 
a position which has been chosen to give a cleanly resolved pattern.
The Cherenkov cones which are internally reflected in the quartz radiator
result in hyperbola-like patterns at the MCP-PMT plane; 
reflections off module sides result in a folding of this pattern.
Chromatic dispersion spreads the lines into bands. Since the quartz is read out by only a single MCP-PMT, the
full pattern is only sampled, which accounts for the  observed  discontinuities.

For the timing measurement, a simultaneous correction is made for time-walk
and integral non-linearities of the NINO/HPTDC electronics
using a data-driven method\,\cite{Brook-2018}. For each single 4-wide pixel row, 
the measured MCP time-stamp for each cluster is plotted relative to the downstream 
borosilicate station (T2) versus the measured 64-wide column position.  An example data distribution is \\

\begin{figure}
\centering
\hspace{0.5truecm} 
\includegraphics[width=0.49\linewidth]{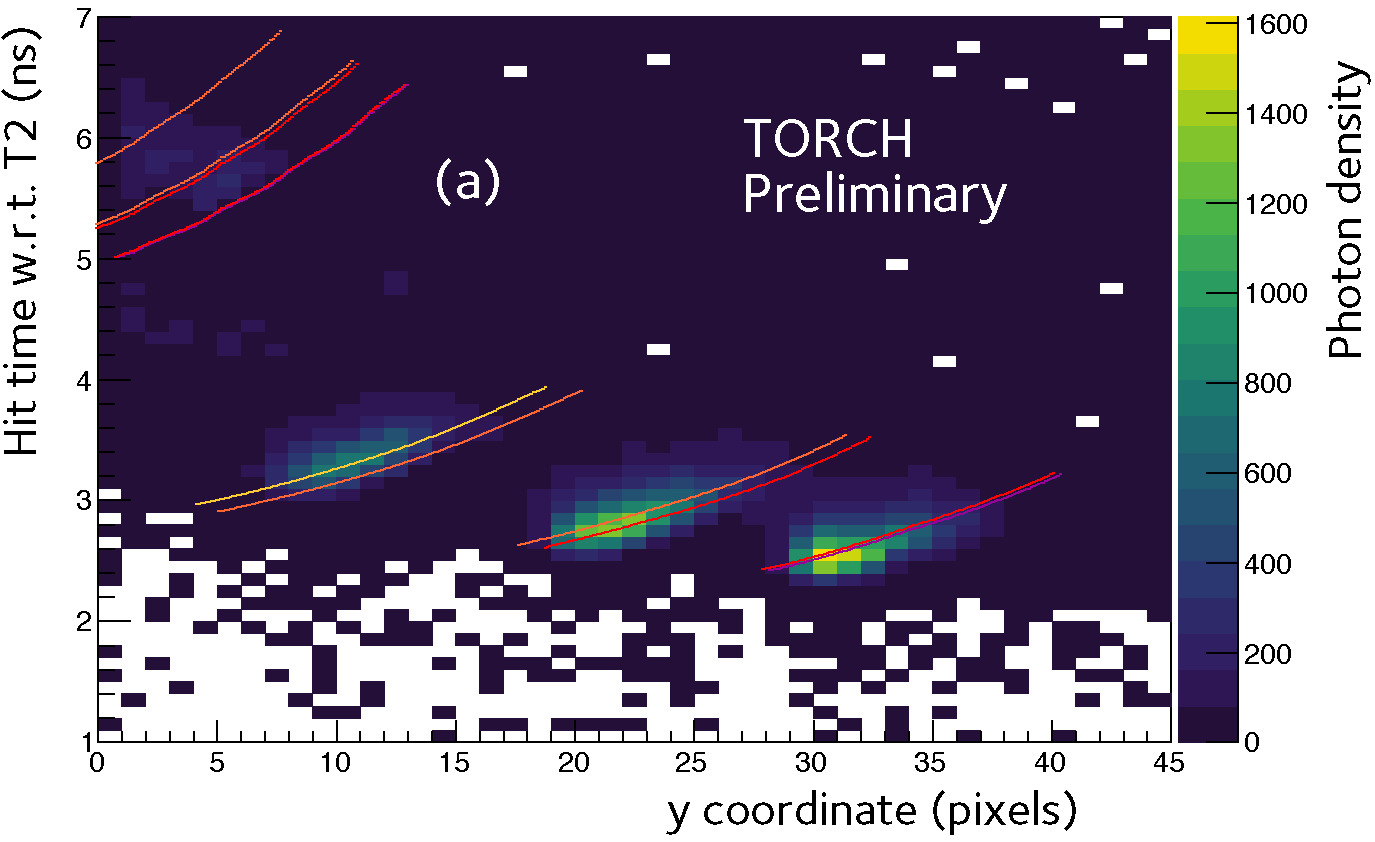}
\includegraphics[width=0.45\linewidth]{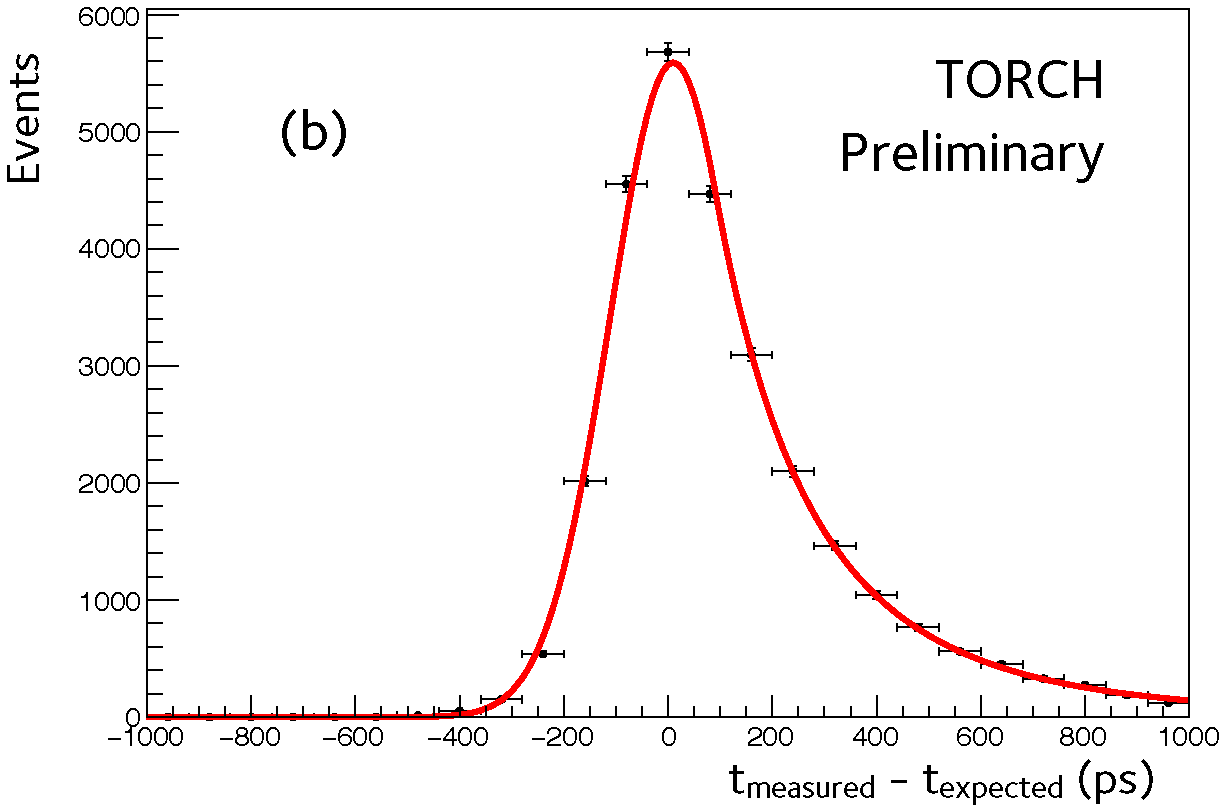}
\caption{(a) The time-of-arrival of single Cherenkov photons from a 5\,GeV/$c$ pion beam,  
relative to the T2 beam time-reference station,
as a function of detected 64-wide column pixel number. 
The overlaid lines 
represent the simulated patterns for light reflected  only off the front and back  faces
of the radiator plate  (purple),  light  
undergoing one (red), two (orange) and three (yellow) reflections off the side faces. The top left distributions
correspond to multiple reflections from the bottom horizontal  face.
(b) The residuals between observed and simulated Cherenkov photon arrival times for 
pixel row 4 (November 2017 data).}
\label{fig:timeprojection}
\end{figure}

\noindent shown in Fig.\,\ref{fig:timeprojection}\,(a) showing good agreement when compared to simulation.
The distribution of
residuals between the measured and simulated times of arrival is shown in  Fig.\,\ref{fig:timeprojection}\,(b).
Core distributions have resolutions (sigmas) of approximately  100 - 125\,ps (which is photon energy and MCP-PMT row dependent).
The tails are due to imperfect calibration and backscattering from the MCP top surface.
The timing resolution of the timing reference is $\sim$50\,ps and, subtracted in quadrature, gives 
$\sim$85 - 115\,ps time resolution of the TORCH demonstrator, 
approaching the target resolution of 70\,ps per photon.
Future improvements are possible such as incorporating 
charge to width calibrations of the front-end electronics and reducing the current limitation imposed by the
100\,ps time binning of the HPTDC.

\section{Development of half-length TORCH prototype}

A  prototype of a half-length TORCH module, 125 $\times$ 66 $\times$ 1\,cm$^3$  (length, width and thickness),  is 
currently under construction. The module will be equipped with ten MCP-PMTs ($\sim$5000 channels).
The radiator plate and focussing block were both procured from Nikon, Japan.
As an incremental step, the small-scale TORCH demonstrator was
equipped with a single 64 $\times$ 8 pixel MCP-PMT and upgraded electronics
and tested in a 5\,GeV/$c$ pion beam in June 2018.   
The pattern of measured  MCP-PMT hits is shown  in Fig.\,\ref{fig:hits}\,(b).
Results are currently being analysed, and calibrations and timing measurements are in progress.
The full-scale module is planned for test-beam running in October/November 2018.

\newpage

\section{TORCH for the LHCb Upgrade II}

The RICH system currently provides PID
for the LHCb experiment\,\cite{ref:LHCb-detector-paper},
where discrimination of pions, kaons and protons  is essential for CP violation measurements, 
exotic spectroscopy and particle tagging.   
However, LHCb has no positive kaon or proton identification below $\sim$10\,GeV/$c$. Therefore 
the proposal is to install TORCH immediately upstream of the RICH\,2 detector, where it would be 
located $\sim$9.5\,m from the proton-proton interaction region\,\cite{LHCb_PhaseII_upgrade}.
Here the total area of TORCH would be 5 $\times$ 6\,m$^2$,  divided into 18 modules, 
each 66\,cm wide and 2.5\,m high with 11 MCP-PMTs per module.

Studies are underway  to evaluate the performance of TORCH in the LHCb experiment. 
A luminosity of 2 $\times$ 10$^{33}$ cm$^{-2}$s$^{-1}$ has been simulated
($\times$\,5 the current  LHCb luminosity), including collision pile-up.
The track entry point in the quartz radiator of TORCH  is  provided
by position information from the  tracking system of the LHCb
spectrometer.  The expected hit distributions 
for the $\pi /K /p$  hypotheses are 
compared to the  MCP-PMT photon spatial hits and arrival times, and log-likelihoods are then computed.

\begin{figure}[h]
\centering 
\includegraphics[width=0.49\columnwidth]{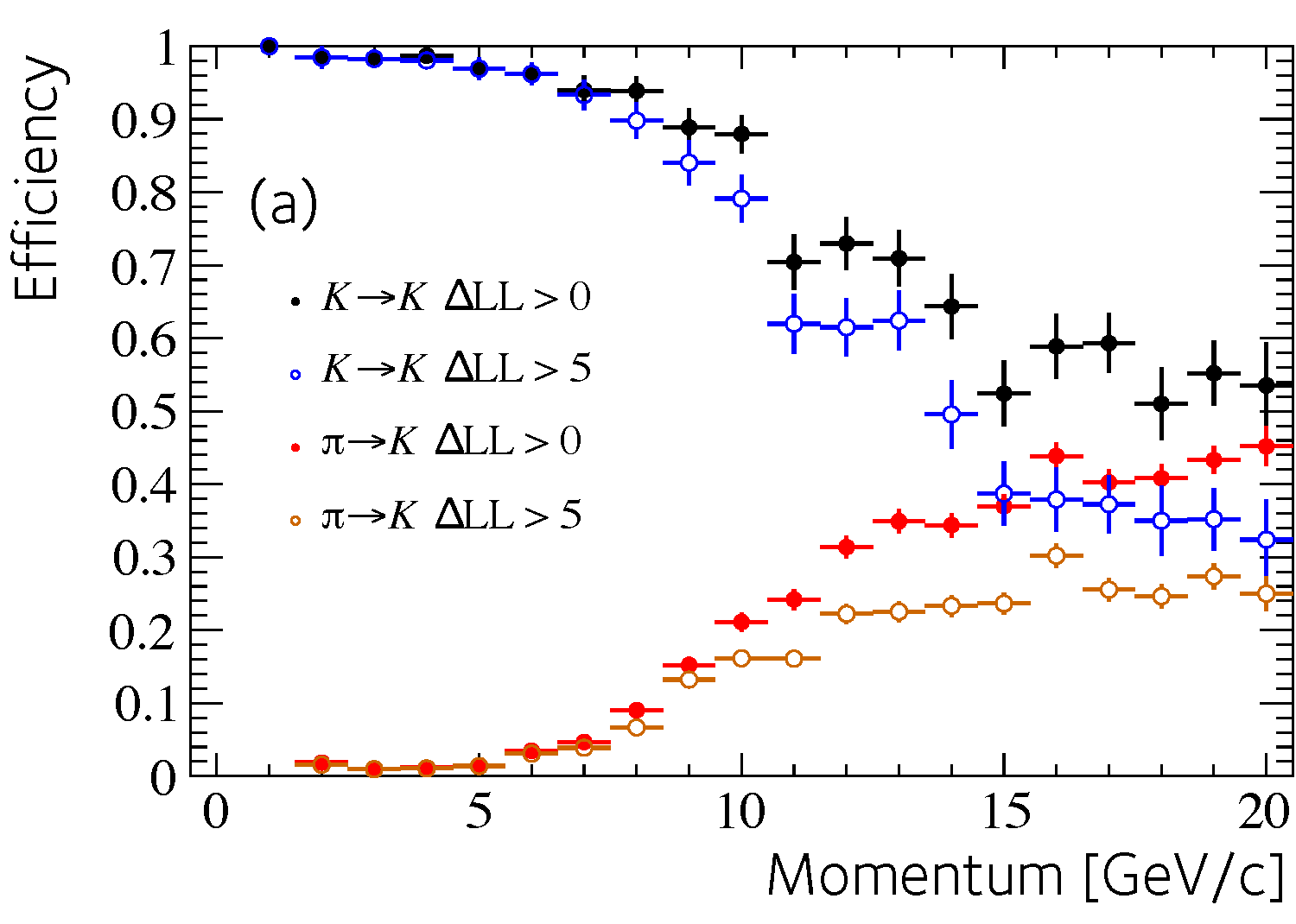}
\includegraphics[width=0.49\columnwidth]{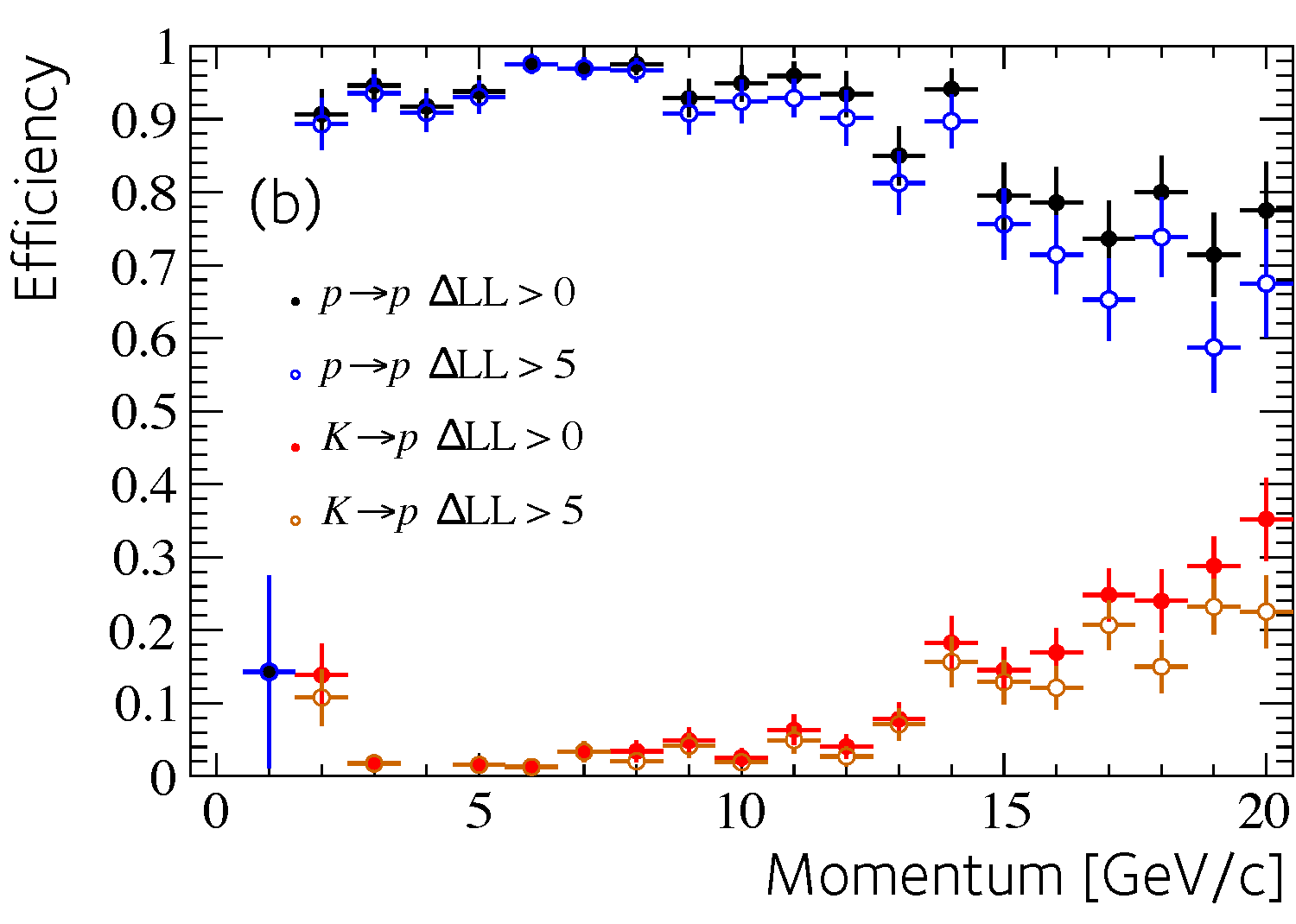}
\caption{The efficiency of TORCH in LHCb to positively identify (a) kaons 
and (b) protons  as a function of momentum and the probability that 
they are misidentied. The curves are for two different delta-log-likelihood cuts and for a luminosity of 
2 $\times$ 10$^{33}$ cm$^{-2}$s$^{-1}$. 
The simulated sample is for heavy flavour decays  in $pp$ collisions,
including pile-up.  }
\label{fig:LHCBefficiency}
\end{figure}

Figure\,\ref{fig:LHCBefficiency}  shows the efficiency of TORCH to positively identify kaons 
and protons as a function of momentum and the probability 
of misidentification. Good separation between $\pi$/K/p in the 2$-$10\,GeV/$c$ range and beyond is observed. 
Studies have also started on key physics channels and tagging performance,
and these  will form the basis of a Technical Proposal to construct a full-scale TORCH detector 
for the start-up of LHC Run\,4, with installation in the Long Shutdown\,3 (LS3) in 2024.

\vspace{0.5truecm}

\newpage

\section{Summary}

The performance of a small-scale TORCH demonstrator  in a 5\,GeV/$c$  mixed pion/proton beam has been reported.
A customised 64 $\times$ 4 pixelated MCP-PMT has been prototyped,
and   timing resolutions of 85--115\,ps per photon
have been measured. With future  improvements, it is hoped to achieve the desired per-photon resolution of 70\,ps.
A half-length TORCH module incorportating  new optics is  under construction, incorporating 
final  64 $\times$ 8 pixelated MCP-PMTs and a  new generation of electronics.
Studies are underway to prepare a  Technical Proposal for TORCH to
be incorporated in the Upgrade-II of the LHCb experiment.

\section*{Acknowledgments}

The support is acknowledged of the Science and Technology Research Council, UK, and of the 
European Research Council  through an FP7
Advanced Grant  (ERC-2011-AdG 299175-TORCH).


\end{document}